\documentclass{aa}  

\usepackage{graphicx}
\usepackage{txfonts}
 \usepackage{threeparttable}
\usepackage{xcolor}

\begin{document}

   \title{Improved thermonuclear rate of $^{42}$Ti($p$,$\gamma$)$^{43}$V and its astrophysical implication in rp-process}
   
 %  \subtitle{New $^{42}$Ti($p$,$\gamma$)$^{43}$V thermonuclear rate}

   \author{S.Q.~Hou\inst{1,2,3},
          C.~Iliadis\inst{4,5},
          M.~Pignatari\inst{6,7,8,9,3},
          J.B.~Liu\inst{1,2},
          T. C. L. ~Trueman\inst{6,7,8,3},
          J.G.~Li\inst{1,2},
          \and
          X.X. Xu\inst{1,2}
          }
    
\authorrunning{S.Q.~Hou et al.} 

  \institute{Key Laboratory of High Precision Nuclear Spectroscopy, Institute of Modern Physics, Chinese Academy of Sciences, Lanzhou 730000, PR China\\
        \email{sqhou@impcas.ac.cn}\\
        \and School of Nuclear Science and Technology, University of Chinese Academy of Sciences, Beijing 100049, PR China\\
        \and NuGrid Collaboration, http://www.nugridstars.org \\
        \and Department of Physics \& Astronomy, University of North Carolina at Chapel Hill, North Carolina 27599-3255, USA \\
        \and Triangle Universities Nuclear Laboratory (TUNL), Duke University, Durham, North Carolina 27708, USA\\
        \and Konkoly Observatory, Research Centre for Astronomy and Earth Sciences, Hungarian Academy of Sciences, H-1121 Budapest, Hungary \\
        \and CSFK, MTA Centre of Excellence, Budapest, Konkoly Thege Mikl\'{o}s \'{u}t 15-17, H-1121, Hungary \\
        \and E. A. Milne Centre for Astrophysics, University of Hull, Kingston upon Hull HU6 7RX, United Kingdom \\
        \and Joint Institute for Nuclear Astrophysics, Center for the Evolution of the Elements, Michigan State University, East Lansing, Michigan 48824, USA}
   \date{Received September 15, 1996; accepted March 16, 1997}

% \abstract{}{}{}{}{} 
% 5 {} token are mandatory
 
  \abstract
   {Accurate $^{42}$Ti($p$,$\gamma$)$^{43}$V reaction rates are crucial for understanding the nucleosynthesis path of the rapid capture process (rp-process) that occurs in X-ray bursts.}
   {We aim to improve the thermonuclear rates of $^{42}$Ti($p$,$\gamma$)$^{43}$V based on more complete resonance information and more accurate direct component, together with the recently released nuclear masses data. Meanwhile, the impact of the newly obtained rates on the rp-process is explored.}
   {We reevaluated the reaction rate of $^{42}$Ti($p$,$\gamma$)$^{43}$V by the sum of the isolated resonance contribution instead of the Hauser-Feshbach statistical model. Meanwhile, a Monte Carlo method is used to derive the associated uncertainties of new rates. The nucleosynthesis simulations are performed via the NuGrid post-processing code ppn.
   }
   {The new %ly obtained 
   rates differ from previous estimations because of using a series of updated resonance parameters and direct S-factor. Compared with the previous results from Hauser-Feshbach statistical model, which assumes compound nucleus $^{43}$V with a sufficiently high-level density in the energy region of astrophysical interest, huge differences exist over the entire temperature region of rp-process interest, even up to 4 orders of magnitude. We consistently calculate the photodisintegration rate using our new nuclear masses via the detailed balance principle and find the discrepancies among the different reverse rates are much larger than the case for the forward rate, up to 10 orders at the temperature of 10$^8$ K. Using a trajectory with a peak temperature of 1.95$\times$10$^9$ K, we perform the 
   rp-process nucleosynthesis simulations to investigate the impact of the new rates. %, and 
   Our calculations show that the adoption of the new forward and reverse rates %can 
   result in abundance variations for Sc and Ca by 128\% and 49\% respectively
   compared to the case using statistical model rates. On the other hand, the overall abundance pattern is not significantly affected.
   The results of using new rates also confirm that the rp-process path does not bypass %nucleus 
   the isotope $^{43}$V.}
   {It is found that the Hauser-Feshbach statistical model is inappropriate to the reaction rate evaluation for $^{42}$Ti($p$,$\gamma$)$^{43}$V. The adoption of the new rates confirms the reaction path of $^{42}$Ti($p$,$\gamma$)$^{43}$V($p$,$\gamma$)$^{44}$Cr($\beta^+$)$^{44}$V is a key branch of the rp-process in X-ray bursts. }


    \maketitle
%-------------------------------------------------------------------

\section{Introduction}
   As the most frequent thermonuclear explosions in the galaxy, Type I X-ray bursts
take place on the surface of a neutron star that accretes matter from a nearby companion star. It is powered by unstable thermonuclear burning of freshly accreted hydrogen and helium material, where three nuclear burning patterns of triple-$\alpha$ reaction, rapid proton-capture ($rp$) process, and $\alpha$-capture proton-emission ($\alpha$$p$)-process are involved (Taam et al. \citeyear{Taam93}; Woosley et al. \citeyear{Woosley_2004}; Fisker et al. \citeyear{Fisker_2008}; Jos\'{e} et al. \citeyear{Jordi10}). Therein, the rp-process can approach the proton drip-line far from the valley of stability via consecutive proton capture on seed nuclei despite in some cases the proton capture has to wait for a $\beta$$^+$- decay before continuing.  Finally, heavier elements with atomic mass number A=60-100 can be synthesized (Wallace \& Woosley et al. \citeyear{Wallace81}; Schatz et al. \citeyear{Schatz98}; Koike et al. \citeyear{Koike99}; Schatz et al. \citeyear{Schatz01}; Koike et al. \citeyear{Koike04}). 

During the rp-process, a large number of short-lived and neutron-deficient nuclei are involved because the reaction path of the rp-process is far away from the valley of stability. Along the rp-process path, the unstable nuclei at which the proton capture process will fiercely compete with the $\beta$$^+$-decay are called the rp-process branching nuclei. $^{42}$Ti is a typical rp-process branching nucleus, from which a splitting of the rp-process nucleosynthesis path will be created. The branching ratio of the proton capture and the $\beta$$^+$-decay for $^{42}$Ti depends on the quantity of the net proton capture flow through $^{42}$Ti, which is determined by the competition between the $^{42}$Ti($p$,$\gamma$)$^{43}$V reaction and its reverse process $^{43}$V($\gamma$,$p$)$^{42}$Ti. Thus, both the accurate forward and reverse reaction rates of $^{42}$Ti($p$,$\gamma$)$^{43}$V are very important for the study of the reaction path in the rp-process.

From Schatz \citeyear{Schatz06}, we know that the forward and reverse reaction rates can be converted mutually via an exp(-Q / kT) involved term, which means the reverse reaction can be directly obtained if the forward reaction rate and reaction Q-value are available. %In light of the reaction $Q$-value of the $^{42}$Ti($p$,$\gamma$)$^{43}$V less than 200 keV, 
Since the $^{42}$Ti($p$,$\gamma$)$^{43}$V reaction $Q$-value is less than 200 keV,
the role of photodisintegration reaction $^{43}$V($\gamma$,$p$)$^{42}$Ti played in the rp-process will be very sensitive to the uncertainties of nuclear masses of the nuclei involved. % in this reaction because 
Indeed, the photodisintegration can more efficiently hinder the proton capture for the reaction with small Q-value relative to that with large Q-value. In addition, %we also know that 
the forward reaction rate of $^{42}$Ti($p$,$\gamma$)$^{43}$V is substantially influenced by the nuclear masses of the involved nuclei: %since 
the alteration of nuclear masses can lead to the change of the resonance energy, which sensitively affects the resonant reaction rate. Therefore, in order to have a complete and comprehensive understanding of the role of $^{42}$Ti($p$,$\gamma$)$^{43}$V reaction in the rp-process, it is worthwhile to make a detailed investigation to its forward and reverse reaction rates.

Until now, the $^{42}$Ti($p$,$\gamma$)$^{43}$V reaction rate has been studied in several previous works (Wormer et al. \citeyear{Wormer94}; Herndl \citeyear{Herndl95}; He et al. \citeyear{He14}; Rauscher \& Thielemann \citeyear{Rauscher00}; Cyburt et al. \citeyear{Cyburt10}). All of them can be categorized into two types according to the way %manner of obtaining 
the final reaction rates are obtained: one is that the rate is thought to be the sum of the rate contribution from each isolated resonance separately; the other is that the reaction rate is obtained based on the Hauser-Feshbach statistical model, which is appropriate only for the case that nuclear level density in the compound nucleus is sufficiently high to make resonances completely overlap. Among them, the works from Wormer et al. \citeyear{Wormer94}; Herndl \citeyear{Herndl95};  He et al. \citeyear{He14} belong to the first type. The earliest study of this reaction is given by \citet{Wormer94} where they used a proton separation energy of 88 keV from \cite{Wapstra83} and consider only four excited states closest to the proton threshold, based on the assumption of $^{43}$V holding exactly the same level structure as its mirror nuclide $^{43}$Ca. One year later, \cite{Herndl95} reevaluated this rate using the different energy levels of $^{43}$V predicted from large-basis shell model calculations instead of assuming the same excitation energies as its mirror states. Furthermore, they consider fewer resonance states, only including the first and second excited states. In \citet{He14}, this reaction rate was updated using an 83 keV proton separation energy obtained from the nuclear masses from AME2012 (Wang et al. \citeyear{wang12}), and for the first time consider the contribution from the 7/2$^+$ resonance at 2.067 MeV besides the first and second excited states. For the reaction rates obtained by the NON-SMOKER Hauser-Feshbach statistical model, there currently exist two versions collected in JINA REACLIB, which are presented by Rauscher et al. in \citeyear{Rauscher00} and Cyburt et al. in \citeyear{Cyburt10}, respectively. In their own evaluation, the proton separation energies of $^{43}$V are thought to be -0.0189 MeV and -0.411 MeV, respectively. As introduced above, the statistical model is only suited for the case of the high-level densities in the compound nuclei (Woosley et al. \citeyear{Woosley78}; Cowan et al. \citeyear{Cowan91}). However, we know that the energy level density of $^{43}$V near the proton threshold is low. Therefore, the results from the statistical model probably carry large uncertainties in calculating the reaction rate of $^{42}$Ti($p$,$\gamma$)$^{43}$V. In this sense, the approach that the reaction rate is computed to be the sum of the individual resonance contribution seems more appropriate for the rate estimation of $^{42}$Ti($p$,$\gamma$)$^{43}$V. %In the case of choosing this manner to evaluate the rate of the $^{42}$Ti($p$,$\gamma$)$^{43}$V, 
By using this approach to calculate the $^{42}$Ti($p$,$\gamma$)$^{43}$V rate,
the major %rate uncertainties come from 
sources of error are the uncertainties of proton separation energy and the level energy of the $^{43}$V excited states.
%Recently released 
A recent evaluation of atomic mass (AME2020) presents the newest nuclear masses data from experiments and theoretical evaluation (Wang et al. \citeyear{Wang21}), from which the new proton separation energy ($S_p$) of $^{43}$V is determined to be 105 keV, with an uncertainty of 40 keV, which is the $S_p$ of highest precision until now. In previous works (Wormer et al. \citeyear{Wormer94}; Herndl \citeyear{Herndl95}; He et al. \citeyear{He14}), some resonant states which have appreciable contributions to the final reaction rate of the temperature larger than 1.1 $\times10^{9}$K were not included. In this work, using the newest nuclear masses and including those resonant states neglected previously, we recalculate the reaction rate of $^{42}$Ti($p$,$\gamma$)$^{43}$V and explore its impact on the rp-process %via a post-processing 
nucleosynthesis %code 
for X-ray burst trajectories extracted from \cite{Schatz01} with a peak temperature of 1.95 $\times10^{9}$K. 

The paper is organized as follows. Sect. \ref{Sect_rate} introduces the basic formalism for astrophysical reaction rate calculations and the new forward and reverse reaction rates for $^{42}$Ti($p$,$\gamma$)$^{43}$V are derived using new nuclear data. We investigate the impact of the new rates on the rp-process by virtue of a post-processing code in Sect. \ref{Sect_Astro}. The conclusions are discussed in Sect. \ref{Sect_Conclu}.

   % Although no hydrodynamical study has been available many workers
   % conjectured that a collapse or rapid contraction will ensue
   % after accumulating the critical mass. The main motivation for
   % this article
   % is to investigate the stability of the static envelope at the
   % critical mass. With this aim the local, linear stability of static
   % radiative gas  spheres is investigated on the basis of Baker's
   % (\citeyear{baker}) standard one-zone model. 

   % Phenomena similar to the ones described above for giant planet
   % formation have been found in hydrodynamical models concerning
   % star formation where protostellar cores explode
   % (Tscharnuter \citeyear{tscharnuter}, Balluch \citeyear{balluch}),
   % whereas earlier studies found quasi-steady collapse flows. The
   % similarities in the (micro)physics, i.e., constitutive relations of
   % protostellar cores and protogiant planets serve as a further
   % motivation for this study.

%--------------------------------------------------------------------

\section{ $^{42}$Ti($p$,$\gamma$)$^{43}$V reaction rate}\label{Sect_rate}
\subsection{The resonant reaction rate}

%-------------------------------------- Two column figure (place early!)

The astrophysical reaction rate consists of a resonant contribution and non-resonant contribution, while the latter term consists of the contributions from direct reaction and subthreshold resonance. For the reaction of $^{42}$Ti($p$,$\gamma$)$^{43}$V, it is confirmed that no subthreshold resonances exist since the first excited state of $^{43}$V from all theoretical prediction remains much higher than the newest proton threshold of 105($\pm40$) keV. For the proton capture reactions in the rp-process, it is often thought that they proceed via narrow and isolated resonances (Timofeyuk et al. \citeyear{Timofeyuk06}). From the classical textbook of nuclear astrophysics, it is known that the reaction rate for a single narrow resonance is expressed as
\begin{equation}
\label{eq1}
N_A\left\langle\sigma v\right\rangle=1.54\times10^{11}\times10^4(\mu T_9)^{-3/2}\omega\gamma \\
\times exp(-\frac{11.605E_r}{T_9}).
\end{equation}
Here, $\mu$ refers to the reduced mass of the colliding system in atomic units. With the same units as the resonance energy $E_r$ (in unit of MeV), the resonance strength $\omega\gamma$ is defined as

\begin{equation}
\label{eq2}
\omega\gamma= \frac{(2J+1)}{(2J_T+1)(2J_p+1)}\frac{\Gamma_p\Gamma_{\gamma}}{\Gamma},
\end{equation}
where $J_p, J_T, J$ denote the spin of the proton, target nucleus, and the resonant state, and $\Gamma_p, \Gamma_{\gamma}$ are the partial width of the entrance channel and exit channel, respectively. The total width $\Gamma$ can be approximated as the sum of $\Gamma_p$ and $\Gamma_{\gamma}$, since other channels are closed in the energy range of our study. Here, the proton width $\Gamma_p$ can be obtained by the following formula:

\begin{equation}
\label{eq3}
 \Gamma_p= \frac{3\hbar^2}{AR^2}P_l(E,R)C^2S_p,
\end{equation}
where $R=1.25(1+42^{1/3})$ fm is the nuclear channel radius, $P_l(E,R)$ is the Coulomb penetrability which can be calculated numerically, and $C^2S_p$ is the spectroscopic factor of a particular state. Both $C^2S_p$ and $\Gamma_{\gamma}$ can be supplied via shell model calculation for the case that no experimental information is available. In the case of several narrow and isolated resonances dominating the cross sections, the total reaction rate is equal to the sum of the individual contributions from every single resonance.

As seen from Eq.(\ref{eq1}), one knows that the reaction rate is determined by two key quantities which are the resonance energy of exponential dependence and the resonance strength of linear dependence, respectively. From the expression of Eq.(\ref{eq3}), we know that the resonance strength is also sensitive to the resonance energy because of its sensitivity to the coulomb penetration factor which dramatically influences the proton width. %Therefore, it becomes very significant to investigate the impact of key nuclear parameters associated with the resonance energy and the resonance strength on the reaction rate. For proton capture reaction, the resonance energy is defined as a quantity that the excitation energy of the compound nucleus state subtracts the threshold energy of proton emission. As seen in eq.3, it also can substantially affect the resonance strength by its sensitive influence in the partial width of different particle decay.
%\subsection{resonance energy}
For proton capture reactions, the resonance energy refers to the energy difference between the excitation energy of the compound nucleus state and the proton separation energy. As mentioned in the introduction, because of adopting the more accurate mass excesses of $^{42}$Ti and $^{43}$V in AME2020, the new value of $S_p$ is fixed to be 105(40) keV, which is about 21.5 keV larger than that used in \cite{He14}. More importantly, this new value has the smallest error up to now. Another key factor determining the resonance energy is the excitation energy of near-threshold states in $^{43}$V. In the current situation, the accurate excitation energy of the $^{43}$V excited states remains unclear because of the vast difficulties to conduct such experiments for short-lived nucleus $^{43}$V. Therefore, one has to resort to other avenues like a theoretical prediction.

At present, three different values for the peak temperatures reached during X-ray bursts have been proposed in the literature (Koike et al. \citeyear{Koike04}; Fisker et al. \citeyear{Fisker_2008}; Schatz et al. \citeyear{Schatz01}), approximately ranging from 1.0 to 1.95 $\times10^{9}$K. According to the estimation of the astrophysical effective energy region, all the excited states of $^{43}$V with excitation energies below 2.17 MeV should be included provided that the peak temperature is set to be 1.95 $\times10^{9}$K. However, the majority of these states below 2.17 MeV are not included in previous works because they think the first and second excited states dominate the reaction rate. Actually, for the case of the peak temperature up to 1.95 $\times10^{9}$K (Schazt et al. \citeyear{Schatz01}), every resonant state, which has a noticeable contribution to the $^{42}$Ti($p$,$\gamma$)$^{43}$V reaction rate, should be taken into account. Note that the resonance excited states which need to carry larger than 3 units of orbit angular momentum to populate will be excluded in our consideration, owing to these states usually playing a negligible role in astrophysical processes (Setoodehnia et al. \citeyear{Setoodehnia20}). All the states of interest are listed in the first column of Table \ref{tab1}.

Because the experimental information of the level structure of the short-lived nucleus $^{43}$V  remains very limited, the excitation energies of the first and second excited states of $^{43}$V used in the previous two evaluations (Herndl \citeyear{Herndl95}; He et al. \citeyear{He14}) are both from shell model calculation. However, we also know that the high-precision prediction of level structures for a certain nucleus is still a huge challenge to shell model theory.
In the present work, six more excited states of $^{43}$V are also considered in addition to the first and second excited states. In this case, instead of adopting shell model prediction, we tend to choose the same approach as used in \cite{Wormer94} that the energy level structures for mirror pair $^{43}$V and $^{43}$Ca are assumed to be identical, but assigning 100 keV uncertainties to $^{43}$V energy levels. The reason we make such an assumption is that a plethora of experimental evidence supports the fact that the mirror energy differences of pf-shell nuclei are within 100 keV for low spin states (O’Leary et al. \citeyear{O'Leary97}; Davies et al. \citeyear{Davies13}). Because of the uncertainty of 40 keV of the new proton separation energy, the uncertainty of resonance energy is determined to be 108 keV.

%Using the new resonance energy obtained via the combination of new nuclear masses and the excitation energy mentioned above, we can calculate the new proton width $\Gamma_p$ for the three excited states, as listed in the \textbf{third} column of table.1 The $\Gamma_{\gamma}$ can be estimated from the lifetime of the corresponding mirror state in $^{43}$Ca via the $\Gamma_{\gamma}=\hbar/\tau$, where $\tau$ is the lifetime of the corresponding excited state. The information used for the calculation of the resonance strength can be summarised in Table.1.

From Eq.(\ref{eq3}), we also know that another key quantity in the determination of the proton emission width is the spectroscopic factor $C^2S$. However, the spectroscopic factors of our target states in $^{43}$V remain unknown because of the scarcity of relevant experiments. Luckily, the corresponding $C^2S$ of its mirror states has been studied extensively by experiments, including stripping reaction ($d,p$) and pick-up reaction ($p,d$). For radiative reactions, the spectroscopic factor should correspond to the value determined by the stripping reaction. Unlike adopting the values from the pick-up reaction for the first resonance in \cite{He14}, we here adopt a different path to ascertain the $C^2S$ of each resonance. Specifically, the average values of $C^2S$ from the stripping reaction works (Endt \citeyear{Endt77}; Endt \& Van Der Leun \citeyear{Endt78}; Brown et al. \citeyear{Brown74}; Dorenbusch et al. \citeyear{Dorenbusch66}) are used for states at 0.593, 0.99, 1.394, 1.957, 2.046 MeV, but for resonances at 0.373, 1.931, and 2.067 MeV, we have to rely on the shell model calculation. As to the uncertainties of the spectroscopic factor $C^2S$ of every resonance we considered, we assume a factor varying from 1.6 to 3 for $C^2S$ values from previous stripping reaction works and a factor of 10 for those from our shell model calculations. The detailed information can be seen in the fifth column of Table\ref{tab1}. 

%Luckily, the Gamma width of these two states can been determined with an accuracy at the level of 10$^{-4}$ eV, which are at least four orders of magnitude smaller than their proton width calculated by using new resonance energy on the assumption that their spectroscopic factor are set to 1. Therefore, their resonance strengths are totally determined by the gamma width for these two states in this situation. Given that the spectroscopic factors for some states of $E_x$ $\textgreater$ 1.394 MeV are unknown currently, they will be set to 1.0 in our calculation.

Using the resonance energy determined from new data, in a combination of the new $C^2S$ values of the states we considered, the new proton width $\Gamma_p$ for each considered excited state can be obtained via Eq.(\ref{eq3}), as listed in the seventh column of Table \ref{tab1}. Regarding the gamma widths $\Gamma_{\gamma}$ of these resonance states, we adopt the values estimated from the lifetime of the corresponding mirror state in $^{43}$Ca via the $\Gamma_{\gamma}=\hbar/\tau$, where $\tau$ is the lifetime of the corresponding excited state. The uncertainty of $\Gamma_{\gamma}$ is uniformly assumed to be a factor of 3 which is simply estimated by the formula of $\gamma$ transition width, together with the energy level difference between $^{43}$V and $^{43}$Ca.  All the information used for the calculation of the resonance strength can be summed up in Table \ref{tab1}.

%Below table is the new data based on adopting new C2S and E_r(Sp=0.1046 MeV),
%ooooooooooooooooooooooooooooooooooooooooooooooooooooooooooooooooooooooooooooooooooooooooooooooooooooooooooooooooooooo
\begin{table*}
\centering
\caption{\label{tab1} The parameters of resonant states used in the calculation of $^{42}$Ti($p$,$\gamma$)$^{43}$V resonant reaction rate, including excitation energy, resonance energies, spin and parity, spectroscopic factor, partial width of gamma and proton, resonance strength.}
\begin{tabular}{ccccrcccc}
\hline\hline
$E_x$ (MeV)        &  $E_r$ (MeV)        & $J^{\pi}$ & $l$ & $C^2S\quad  $               & $\Gamma_{\gamma}$(eV)&  $\Gamma_p$ (eV)          & $\omega\gamma$ (eV)        \\
\hline
0.373 (0.100)  &  0.268 (0.108)  & 5/2$^-$   &3    & 0.0015 (10) & 1.30$\times10^{-5}$ (3.0)  &  2.48$\times10^{-13}$ & 7.44$\times10^{-13}$  \\

0.593 (0.100)  &  0.488 (0.108)  & 3/2$^-$   &1    & 0.015 (2.0)  & 5.60$\times10^{-6}$ (3.0)  &  6.50$\times10^{-5}$ & 1.03$\times10^{-5}$   \\

%2.046  &  1.941  & 3/2$^-$   & 0.23\footnotemark[2} & 3.15$\times10^{-4}$  &  2.95$\times10^{3}$   &                 \\
0.990 (0.100)  &  0.885 (0.108)  & 3/2$^+$   &2    & 0.033 (2.0)  & 9.30$\times10^{-6}$ (3.0)  &  2.10$\times10^{-2}$ & 1.86$\times10^{-5}$   \\
%==========================================================================

1.394 (0.100)  &  1.289 (0.108)  & 5/2$^+$   &2    & 0.0042 (3.0) & 2.47$\times10^{-4}$ (3.0)  &  1.40$\times10^{-1}$ & 7.41$\times10^{-4}$   \\

1.931 (0.100)  &  1.826 (0.108)  & 5/2$^-$   &3    & 0.0009 (10)  & 3.93$\times10^{-3}$ (3.0)  &  5.40$\times10^{-2}$ & 1.18$\times10^{-2}$   \\

1.957 (0.100)  &  1.852 (0.108)  & 1/2$^+$   &0    & 0.019  (2.0) & 4.15$\times10^{-4}$ (3.0)  &  3.68$\times10^{+2}$  & 4.15$\times10^{-4}$   \\

2.046 (0.100)  &  1.941 (0.108)  & 3/2$^-$   &1    & 0.24 (1.6)  & 5.70$\times10^{-4}$ (3.0)  &  3.25$\times10^{+3}$  & 1.14$\times10^{-3}$    \\

2.067 (0.100)  &  1.962 (0.108)  & 7/2$^-$   &3    &0.0023 (10)  & 2.20$\times10^{-2}$ (3.0)  &  
2.50$\times10^{-1}$ & 8.09$\times10^{-2}$  \\

\hline
\end{tabular}

\end{table*}

%[31/Sm05]S. M. Smith et al., Nucl. Phys. A 113, 303 (1968).
%[34]G. Brown, A. Denning, and J. G. B. Haigh, Nucl. Phys. A 225, 267 (1974).
%[Yn01] J.L.Yntema ¨C Phys.Rev. 186, 1144 (1969)
%[Co06]T.W.Conlon, B.F.Bayman, E.Kashy ¨C Phys.Rev. 144, 941 (1966)
\subsection{Direct reaction contribution}
  In this work, the direct reaction contribution is recalculated via a method of using a hard sphere scattering potential (Iliadis \& Wiescher \citeyear{Iliadis04}), which can exclude the possibility that the potential resonance contribution created by the particular choice of the scattering potential is mistakenly taken as a part of the direct contribution, as in \cite{He14}. The actual direct S-factor we obtained can be described by a truncated polynomial $S_{dc}(E)=S(0) + S^{\prime}(0)E +S^{\prime\prime}(0)E^2$, where the corresponding polynomial coefficients are set to be $S$=0.0042 MeV$\cdot$b, $S^{\prime}$=0.0039 b, $S^{\prime\prime}$=5.54$\times$10$^{-5}$ b$\cdot$MeV$^{-1}$ via a $\chi^2$-fit. Our value is about ten times smaller than those used in previous works (Herndl \citeyear{Herndl95};  He et al. \citeyear{He14}).
  
\subsection{Total reaction rate}
\begin{figure}[htbp]
\includegraphics[width=\columnwidth]{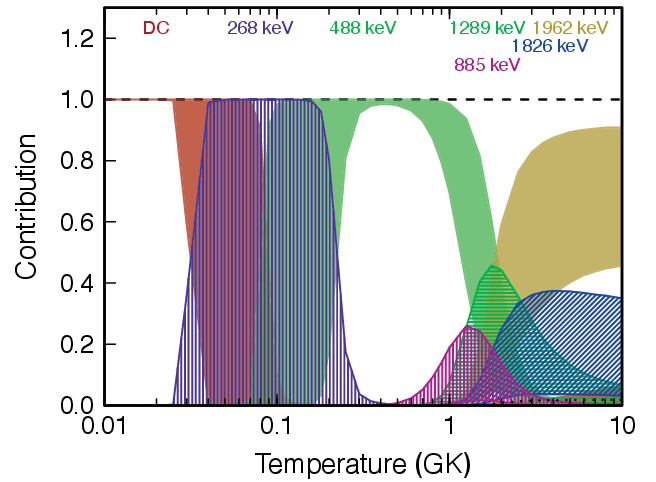}
% \resizebox{8.3cm}{!}{\includegraphics{rates2.eps}}
\caption{\label{fig_1} (Color online) The fractional contributions to the total $^{42}$Ti($p$,$\gamma$)$^{43}$V reaction rate. Resonances are labeled by their center-of-mass resonance energies and the label DC refers to the direct capture process. The contribution ranges are shown as colored bands, with the band thickness representing the uncertainty of the contribution.}
\end{figure}

\begin{figure}[htbp]
\includegraphics[width=\columnwidth]{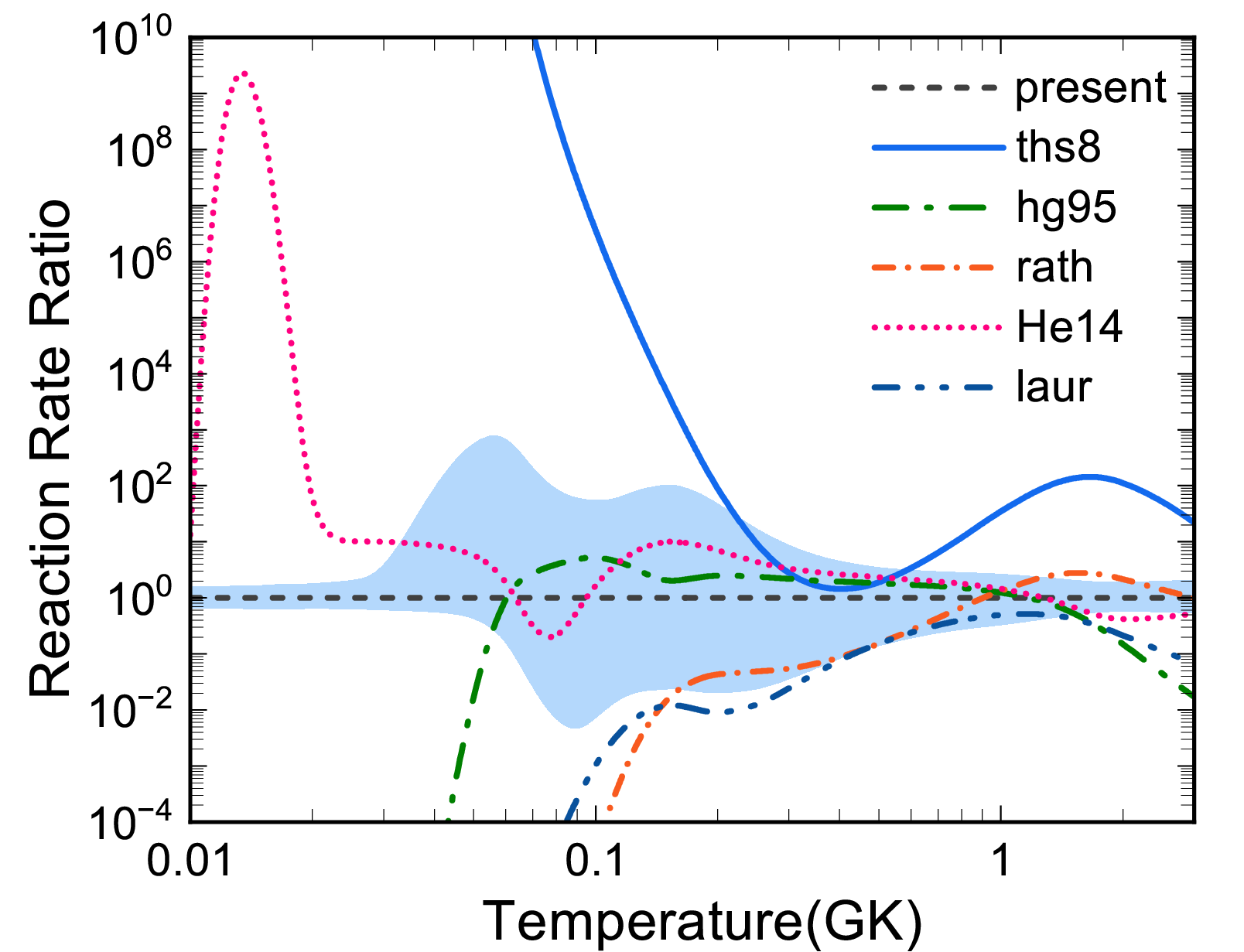}
\caption{\label{fig_2} (Color online) Ratio of previous rates $^{42}$Ti($p$,$\gamma$)$^{43}$V normalized to present recommended rates (the median rates in \ref{tab2}). The different patterns of lines correspond to the rates from Cyburt et al. \citeyear{Cyburt10}; Herndl \citeyear{Herndl95}; Rauscher \& Thielemann \citeyear{Rauscher00};  He et al. \citeyear{He14};  Wormer et al. \citeyear{Wormer94}}, as marked in ths8, hg95, rath, He14, laur, respectively. The shallow blue-shaded areas correspond to 68\% coverage probabilities, as marked Low and High in Table \ref{tab2}.
\end{figure}

Based on the information of resonances and direct reactionintroduced above, the total reaction rate and associated upper and lower limits can be directly obtained by Monte Carlo sampling of all corresponding uncertainties listed in Table \ref{tab1}(Longland et al. \citeyear{Longland10}). The final result is shown in  Fig. \ref{fig_1}. It is clearly seen that the direct reaction and the resonance at E$_r$=268 keV dominate in the temperature range of $T_9$$\leq$0.04 and 0.04$\textless$$T_9$$\textless$0.16 (temperature $T_9$=10$^9$ Kelvin), respectively. The reaction rate over the temperature of 0.16$\textless$$T_9$$\textless$1.6 is totally determined by the E$_r$=488 keV resonant state. The resonances at E$_r$=0.885, 1.289, 1.826, 1.962 MeV have considerable contributions to the rate in the temperature range of $T_9$$\textgreater$1.1, while those of E$_r$=1.852, 1.941 MeV contribute minimally. Fig. \ref{fig_2} plots the new total rate and the associated uncertainties arising from the uncertainties of the resonance energy and the spectroscopic factor. For comparison, the other rates from previous works, including those compiled in JINA REACLIB and \cite{He14}, are also added. It can be seen that our result is consistent with the results from hg95 and He14 over a wide temperature range from 0.2 to 1 $\times10^{9}$K, except for the region of the temperature larger than 1 $\times10^{9}$K and less than 2 $\times10^{8}$K. However, large differences arise if we compare with the results from ths8 and rath - this is due to both of them being obtained by the statistical model. As to the difference between laur's rate and ours, it is mainly caused by different resonance strengths and resonance energies.

\begin{figure}[htbp]
\includegraphics[width=\columnwidth]{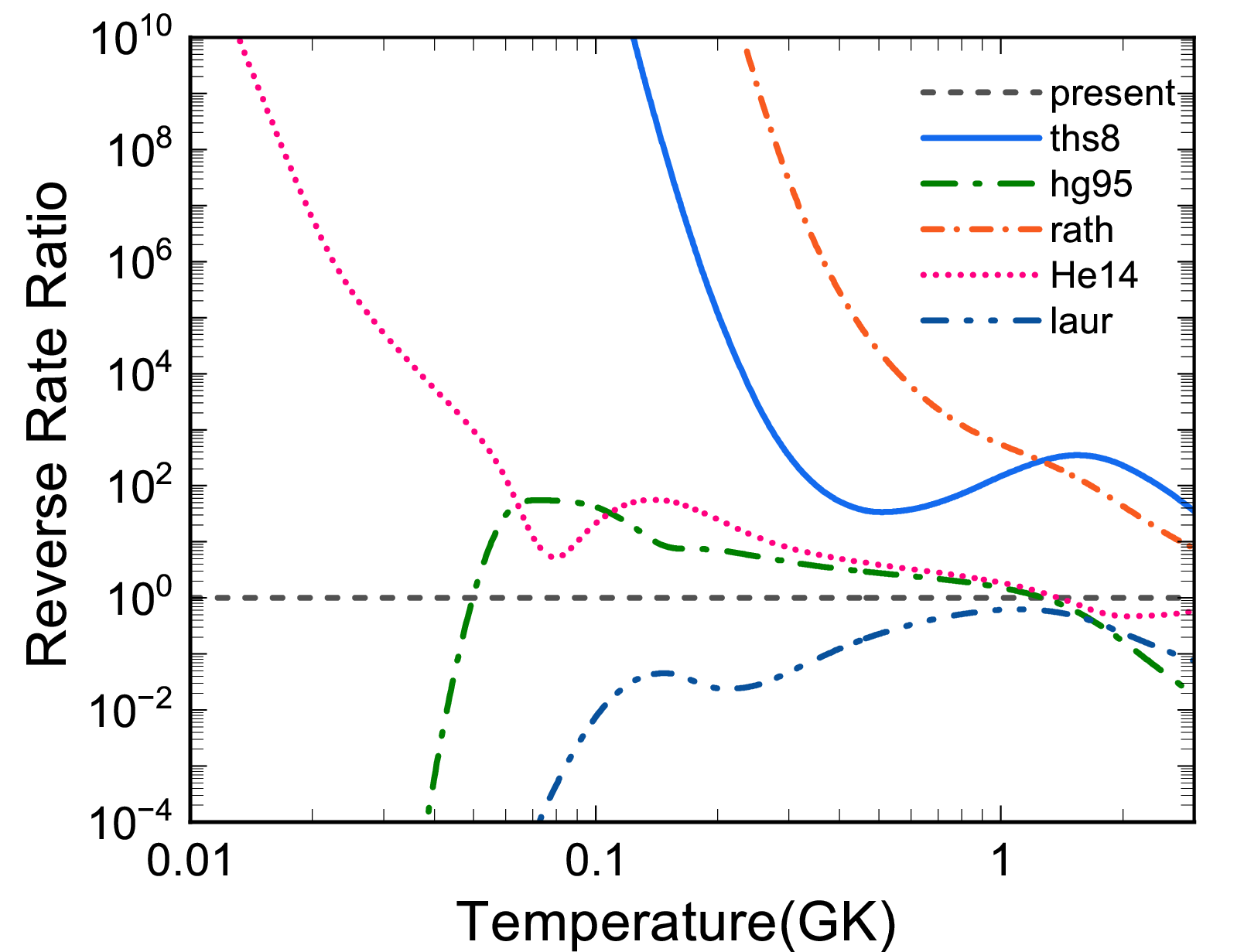}
\caption{\label{fig_3} (Color online) Same as Fig. \ref{fig_2}, but for the reverse rates.}
\end{figure}

The newly obtained reaction rate can be well fitted (less than 2.63\% error in 0.01$\sim$10 $\times10^{9}$K) by the following analytic expression in the standard seven-parameter format of REACLIB:
%\begin{figure*}[h!]
%\centering
%\begin{gather}
%\label{fig9}
$N_A\left\langle\sigma v\right\rangle=
\mathrm{exp}(-281.757+0.13145T_9^{-1}-76.4343T_9^{-1/3}\notag \\+452.566T_9^{1/3}-119.778T_9+15.7643T_9^{5/3}-  85.672\text{ln}(T_9)) \notag \\
+\mathrm{exp}(-176.247-6.23679T_9^{-1}+105.881T_9^{-1/3} \notag \\+71.9063T_9^{1/3} -12.9155T_9 
+0.844572T_9^{5/3}+28.9084\text{ln}(T_9))\notag \\ 
+\mathrm{exp}(213.588+1.51534T_9^{-1}-213.87T_9^{-1/3}-66.3678T_9^{1/3} \notag \\ +79.9314T_9 
-20.7241T_9^{5/3} -89.0426\text{ln}(T_9))
+\mathrm{exp}(-47.6958 \notag \\-1.52387T_9^{-1}-71.8779T_9^{-1/3}+120.168T_9^{1/3} -6.08624T_9 \notag \\
-0.0270441T_9^{5/3} -51.2377\text{ln}(T_9)).$
%\end{gather}
%\end{figure*}
%Here, $T_9$ indicates the temperature in units of 10$^9$ Kelvin. 

\begin{table}[!htb]
	\caption{\label{tab2}The new total reaction rate for $^{42}$Ti($p$,$\gamma$)$^{43}$V}
%	\label{Total_reaction_rate }
		\begin{tabular}{ccccc}
  \hline\hline
			T(GK) & Low & Median & High & f.u.\\
			\hline
		0.01 &  1.75$\times10^{-58}$  & 2.77$\times10^{-58}$  & 4.44$\times10^{-58}$ & 1.599\\
			
		0.02 &  7.33$\times10^{-45}$  & 1.18$\times10^{-44}$  & 2.09$\times10^{-44}$ &7.992\\
			
		0.03 & 2.98$\times10^{-38}$   & 5.11$\times10^{-38}$  & 1.29$\times10^{-37}$&15.11\\
			
		0.04 & 4.61$\times10^{-34}$   & 8.34$\times10^{-34}$  &9.45$\times10^{-32}$&26.77\\
			
		0.05 &  4.70$\times10^{-31}$  & 9.96$\times10^{-31}$  &6.95$\times10^{-28}$&41.76\\
			
		0.06 & 9.55$\times10^{-29}$ & 4.71$\times10^{-28}$  &4.54$\times10^{-25}$ &57.35\\
			
		0.07 &  7.52$\times10^{-27}$  & 3.35$\times10^{-25}$  &6.71$\times10^{-23}$&68.23\\
			
		0.08 &  3.15$\times10^{-25}$  & 5.29$\times10^{-23}$  &4.38$\times10^{-21}$&72.07\\
			
		0.09 &  1.05$\times10^{-23}$  & 2.77$\times10^{-21}$  &1.59$\times10^{-19}$&70.59\\
			
		0.10 &  4.95$\times10^{-22}$  & 7.12$\times10^{-20}$  &3.89$\times10^{-18}$&67.52\\
			
		0.20 &  2.07$\times10^{-13}$  & 1.09$\times10^{-11}$  &5.42$\times10^{-10}$&37.81 \\
			
		0.30 &  1.70$\times10^{-09}$  & 4.98$\times10^{-08}$  &4.25$\times10^{-07}$&14.32\\
			
		0.40 & 2.48$\times10^{-07}$ & 3.07$\times10^{-06}$  &1.38$\times10^{-05}$&7.477 \\
			
		0.50 &  4.67$\times10^{-06}$  & 3.42$\times10^{-05}$  &1.17$\times10^{-04}$&5.334\\
			
		0.60 &  3.04$\times10^{-05}$  & 1.63$\times10^{-04}$  &5.01$\times10^{-04}$&4.447\\
			
		0.70 & 1.12$\times10^{-04}$   & 4.90$\times10^{-04}$  &1.42$\times10^{-03}$&3.976\\
			
		0.80 & 2.95$\times10^{-04}$   & 1.12$\times10^{-03}$  &3.15$\times10^{-03}$&3.648\\
			
		0.90 & 6.25$\times10^{-04}$   & 2.15$\times10^{-03}$  &5.83$\times10^{-03}$&3.369\\
			
		1.00 &   1.15$\times10^{-03}$ & 3.63$\times10^{-03}$  &9.66$\times10^{-03}$&3.111\\
			
		2.00 &  8.54$\times10^{-02}$  & 1.54$\times10^{-01}$  &2.85$\times10^{-01}$&1.865\\
			
		3.00 & 9.29$\times10^{-01}$   & 1.81$\times10^{+00}$ &3.72$\times10^{+00}$ &2.040\\
			
		4.00 & 3.29$\times10^{+00}$   & 6.69$\times10^{+00}$  &1.45$\times10^{+01}$&2.131\\
			
		5.00 & 6.78$\times10^{+00}$   & 1.40$\times10^{+01}$  &3.11$\times10^{+01}$&2.172\\
			
		6.00 &  1.06$\times10^{+01}$  & 2.20$\times10^{+01}$  &4.93$\times10^{+01}$&2.195\\
			
		7.00 &  1.42$\times10^{+01}$  & 2.95$\times10^{+01}$ &6.67$\times10^{+01}$ &2.209\\
			
		8.00 &  1.72$\times10^{+01}$  & 3.60$\times10^{+01}$  &8.14$\times10^{+01}$&2.218\\
			
		9.00 &  1.97$\times10^{+01}$  & 4.10$\times10^{+01}$  &9.36$\times10^{+01}$&2.224\\
			
		10.0  & 2.15$\times10^{+01}$    & 4.49$\times10^{+01}$  &1.02$\times10^{+02}$&2.229\\
   \hline
		\end{tabular}
  \tablefoot{In units of cm$^{3} $mol$^{-1}$ s$^{-1}$. Columns 2, 3, and 4 list the 16th, 50th, and 84th percentiles of the total rate probability density at given temperatures;  f.u. is the factor uncertainty based on Monte Carlo sampling of the total reaction rate. The total number of samples at each temperature was 10 000.}
\end{table}

It is well known that the forward and reverse reaction rates are tightly connected by the detailed balance principle. For the $^{42}$Ti($p$,$\gamma$)$^{43}$V reaction, its reverse reaction rate (also called photodisintegration rate) can be calculated directly by the following expression (Schatz \& Ong \citeyear{Schatz_2017})

\begin{equation}
\label{eq4}
\lambda_{(\gamma,p)}=\frac{2G_f}{G_i}\left(\frac{\mu kT}{2\pi\hbar^2}\right)^{3/2}\mathrm{exp}\left(-\frac{Q_{(p,\gamma)}}{kT}\right)\left\langle\sigma v\right\rangle_{(p,\gamma)}
\end{equation}
where $G_i$ and $G_f$ are the partition functions of initial and final nuclei. The ratio of the ($\gamma,p$) reaction rate to the ($p,\gamma$) rate depends exponentially on $Q_{(p,\gamma)}$ and is therefore very sensitive to nuclear masses. In light of new nuclear masses used in this study, we investigate the impact of the variation of nuclear masses of the nuclei involved in this reaction on the $\lambda_{(\gamma,p)}$ rate. Fig. \ref{fig_3} shows the new $\lambda_{(\gamma,p)}$ rate compared to those from JINA REACLIB and \cite{He14}. It is seen clearly that the two rates from the statistical model (rath and ths8) are much higher than those obtained by the sum of the contributions from narrow and isolated resonances. This large discrepancy is mainly due to their negative reaction Q-values, which are -18.9 keV for ths8 and -411 keV for rath, respectively. Generally speaking, larger $\lambda_{(\gamma,p)}$ rates mean stronger effects of impeding the proton capture on $^{42}$Ti, resulting in a reduction of the net reaction flux pass through $^{42}$Ti($p$,$\gamma$)$^{43}$V reaction.

%\begin{figure}
%\resizebox{8.3cm}{!}{\includegraphics{Rate_Ti42_pg_V43.eps}}
%\caption{\label{fig1} (Color online) The reaction rate of $^{42}$Ti($p$,$\gamma$)$^{43}$V is shown as a function of temperature in unit of GK .}
%\end{figure}

\section{Astrophysical impact for X-ray bursts}\label{Sect_Astro}

Note that the new forward and reverse rates for $^{42}$Ti($p$,$\gamma$)$^{43}$V are remarkably  different from the previous rates in the temperature region of 1 $\times10^{7}$K to 2 $\times10^{9}$K. Therefore, it is worthwhile to explore the impact of the new rates on the rp-process in type I X-ray bursts. We perform the rp-process simulation using the one-zone post-processing nucleosynthesis code ppn, a branch of the NuGrid framework \citep{NuGrid08, Pavel14}. 
%\marco{[I do not understand what you want to say here below. First talk about the trajectory that you use. Then you can compare with other trajectories. See the changes I propose here below.]} {\color{blue} [SQ: From Fig.2 and Fig.3, we know our new rates have no too many differences from the rate by He14 in the temperature range 0.2 - 1.4 GK. In the paper of He14, they also list the abundance variations of isotopes, but different from our results. I compare with K04 trajectory for the purpose of explaining the reason why they are different. Do you mean I could delete the content about the result using K04 trajectory?]}
%Unlike using the K04 profile (\textbf{Koike et al. \citeyear{Koike04}}) in \textbf{\cite{He14}} with a peak temperature of 1.35 GK, 
We choose here a trajectory from \cite{Schatz01}, with a peak temperature %amounting to 
of 1.95 $\times10^{9}$K. %, which is helpful to show the astrophysical impact of the new rates in the temperature range 1.35-1.95 GK. 
The solar abundances from AG89 \citep{AG89} are used as the initial composition of the accreted material from the companion star. Three runs are performed on the condition that all the nuclear physics and model inputs are kept unchanged except the forward and reverse reaction rates for $^{42}$Ti($p$,$\gamma$)$^{43}$V taken from different sources. In the first run, we use our new rates. The ths8 rates are used for the second run and rates from rath %are used 
in the third run. %The reason that we here choose the ths8 and rath rates as the objectives of comparison is due to both the forward rates and reverse rates from them having large differences from our new rates, as shown in Figures \ref{fig_2} and \ref{fig_3}. 
Our rp-process simulations show that the adoption of different rates for the $^{42}$Ti($p$,$\gamma$)$^{43}$V reaction does not have a noticeable influence on the 
sum of the energies released from nuclear reactions (which is indicative of the nuclear energy generation rate), but can result in considerable abundance variations for some isotopes.
Table \ref{tab3} shows a comparison of the calculated abundances of such isotopes for different values of the forward and reverse rates of the $^{42}$Ti($p$,$\gamma$)$^{43}$V reaction. 
The largest variations are obtained for the radioactive isotopes $^{43}$V and $^{44}$Cr, both in the order of magnitude of 4. 

%Notice also that our results differ from the predicted abundances from \cite{He14}, which uses the K04 trajectories (\textbf{Koike et al. \citeyear{Koike04}}). This is due to the much higher temperatures reached by the \textbf{\cite{Schatz01}} trajectory compared to a temperature peak of 1.35 GK obtained by K04.

\begin{figure}
\resizebox{8.3cm}{!}{\includegraphics{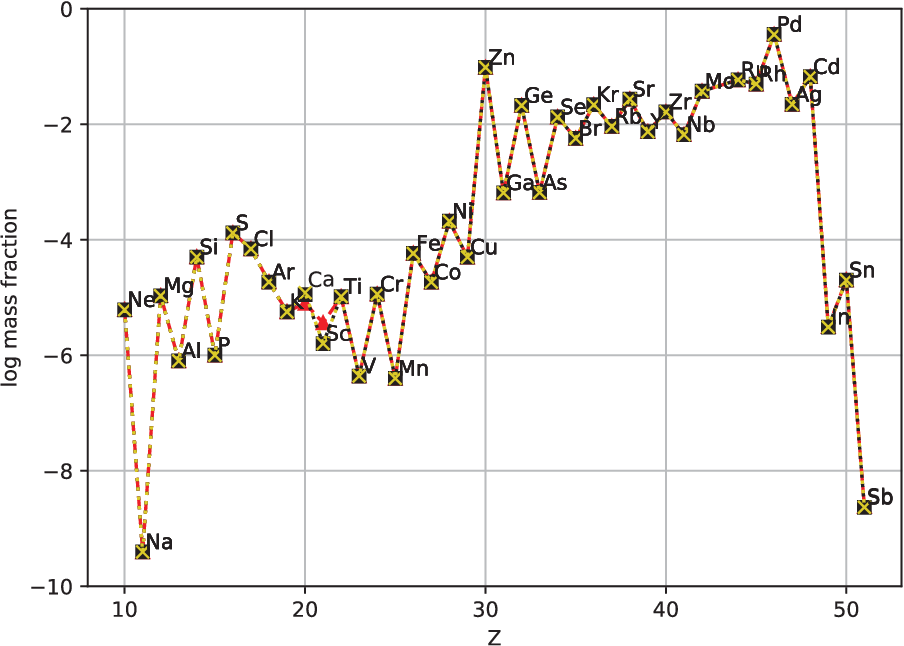}}
\caption{\label{fig_4} (Color online) The decayed elemental abundance distribution in rp-process for cases with different forward and reverse reaction rates for $^{42}$Ti($p$,$\gamma$)$^{43}$V. The red triangle, black square, and yellow cross correspond to cases using new rate, ths8, and rath, respectively.}
\end{figure}

%mainly due to choosing a different profile with a higher peak temperature which can drive the reaction flow toward producing higher Z isotopes. 
To investigate the impact of the variation of $^{42}$Ti($p$,$\gamma$)$^{43}$V reaction rate on the final rp-process composition in the X-ray burst, we calculate the final composition of hydrogen burning ashes. The element abundance after the full decay of all the unstable isotopes is plotted in Fig. \ref{fig_4}. It is found that the abundances of Ca and Sc have a prominent change for the case using our new rate compared with the other two cases using ths8 and rath rates. The decrease of 49\% in Calcium abundance is caused by $^{42}$Ti decay to $^{42}$Ca, while the increase of 128\% in Scandium abundance is due to $^{45}$Ti, $^{45}$V, and $^{45}$Cr decay to $^{45}$Sc. On the other hand, the overall abundance distribution obtained from the calculations is not modified.
Notice that the abundances shown in Figure~\ref{fig_4} differ from the predicted abundances shown by \cite{He14}, which uses the K04 trajectories (Koike et al. \citeyear{Koike04}). This is due to the much higher temperatures reached by the \cite{Schatz01} trajectory compared to a temperature peak of 1.35 $\times10^{9}$K obtained by K04. In order to test the impact of such a relevant difference, we have performed analogous nucleosynthesis calculations also using the K04 trajectory. Similar results previously discussed for the \cite{Schatz01} trajectory are obtained, with the largest variations still in the Ca-Sc region (up to 73\% decrease in Calcium abundance and 200\% increase in Scandium abundance) but with the overall abundance pattern unchanged.

%\marco{[MP: here my main question would be what would have been the results if you would have made the same tests, but using the K04 trajectory. Would you obtain the same results? Can you check? I think so, but it is better to know. Then, I would write a sentence, something like.. we have performed analogous nucleosynthesis calculations also using the K04 trajectory. The same results previously discussed are obtained, with the largest variations in the Ca-Sc region, and with the overall abundance pattern (mostly reaching up to  ... in this case) not modified overall. ]} {\color{blue} [SQ: we have performed analogous nucleosynthesis calculations also using the K04 trajectory. The same results previously discussed are obtained, with the largest variations in the Ca-Sc region, and with the overall abundance pattern (mostly reaching up to... 73\% decrease in Calcium abundance and 200\% increase in Scandium abundance in this case) not modified overall.]}

\begin{figure*}
\centering
\includegraphics[width=16.6cm]{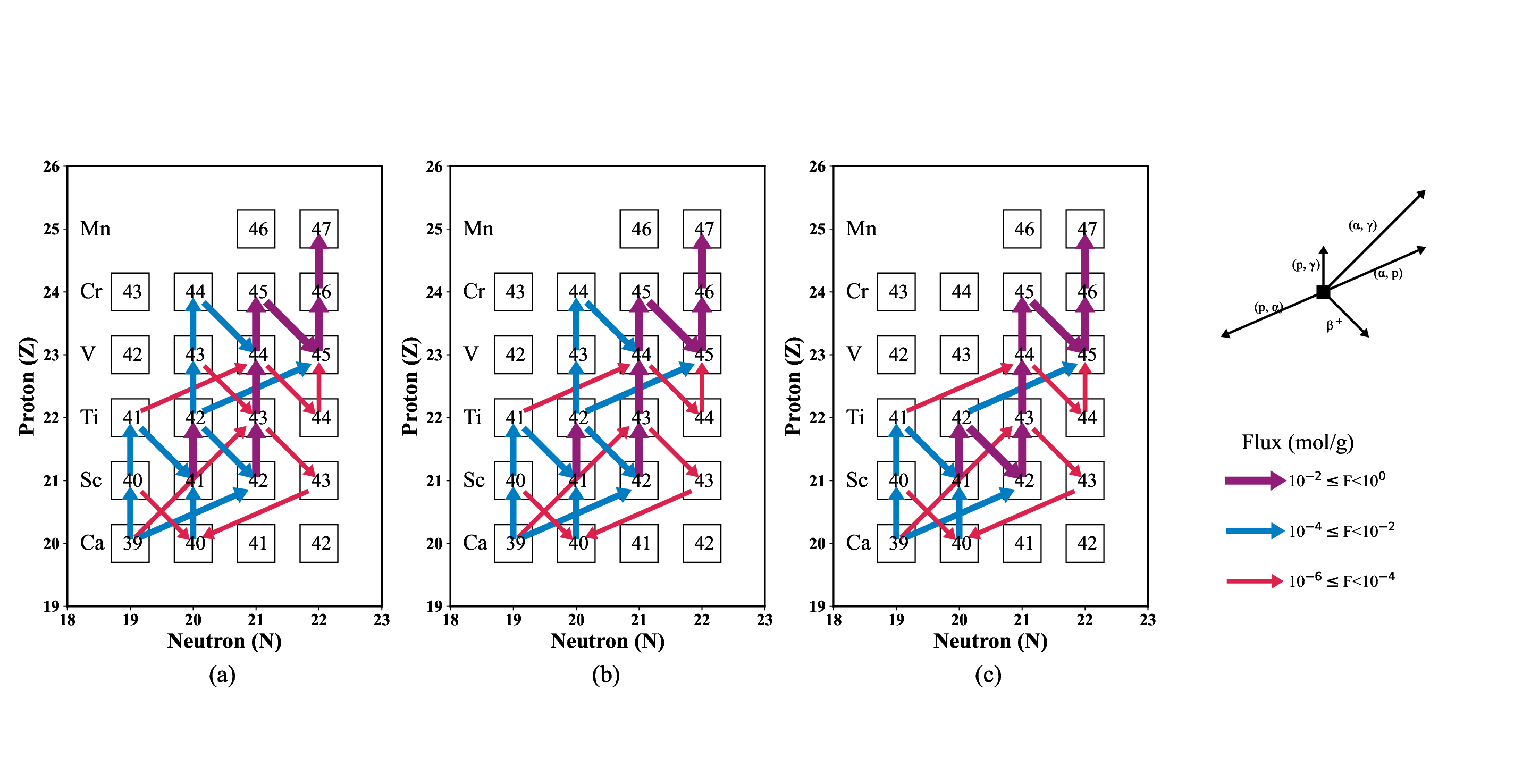}
\caption{\label{fig_5} (Color online) The main reaction flows in rp-process for the adoption of different forward and reverse reaction rates for $^{42}$Ti($p$,$\gamma$)$^{43}$V. Panels (a), (b), (c) are for the cases using new rate, ths8 rate, and rath rate, respectively. The reaction flow is integrated over the entire X-ray burst duration. The thickness of the arrow depicts the magnitude of the reaction flux.}
\end{figure*}

To explore the impact of different rates on the rp-process path, the net reaction flow  between two nuclei $i$ and $f$ is defined as 
\begin{equation}
\label{eq_flux}
\int_t{\dot{Y}_{i}(i\rightarrow f)-\dot{Y}_{f}(f\rightarrow i)}\,dt,
\end{equation}
where, $\dot{Y}_{i}(i\rightarrow f)$ is the partial rate of change of the isotopic abundance ${Y}_{i}$, induced by the particular reaction under consideration that converts the initial nuclide $i$ into final nuclide $f$. Like the definition in \cite{Schatz_2017}, the main reaction flow for an X-ray burst was identified by requiring a net reaction flow integrated over the burst duration either leading to or from the nuclide of at least 10$^{-5}$ mole g$^{-1}$. 
Fig. \ref{fig_5} plots the main reaction flux integrated over the entire rp-process duration. Panel (a) is for the case using the new forward and reverse rate of $^{42}$Ti($p$,$\gamma$)$^{43}$V, and panels (b) and (c) are for ths8 rate and rath rate, respectively. It can be seen that the main rp-process paths are basically same for cases using new rates and ths8 rates. The obvious difference between them is the net flow from $\beta^+$ decay of $^{43}$V  becomes larger when using new rates. This is attributed to a relatively small $^{43}$V($\gamma$,$p$) photodisintegration rate that permit more material through nucleus $^{43}$V, which results in the flow increase of $\beta^+$ decay of $^{43}$V. For panel (c), the nucleosynthesis path varies dramatically. The reaction flow passing $^{42}$Ti($p$,$\gamma$)$^{43}$V is totally prevented by the extremely strong $^{43}$V($\gamma$,$p$)$^{42}$Ti rate, which is due to the reaction Q-value of -411 keV. The path marked by the purple arrow becomes the only channel to guide the reaction flow into the higher Z region. We have seen that in this specific case the overall production of heavier isotopes is not affected, and only a strong impact on the local isotopes is obtained. Nevertheless, it would be different for branching points involving heavier isotopes, especially those with long lifetimes, like $^{64}$Ge and $^{72}$Kr \citep{Schatz06, Schatz_2017}. This is due to the significant lifetime difference between $^{64}$Ge and $^{65}$As can more efficiently regulate the  the progress of the rp-process through the $(p,\gamma$) and $\beta$-decay of $^{64}$Ge. The rate variation of the longer lifetime branching nuclei involved reaction will have an overall effect on the production of heavier isotopes in comparison to those of reactions involving short lifetime branching nuclei like $^{42}$Ti. Therefore, accurate nuclear masses and reaction rates are crucial for a comprehensive understanding of the rapid proton capture in X-ray bursts.

\begin{table}
\centering
\caption{\label{tab3} The calculated abundances for isotopes affected sensitively by three groups of forward and reverse rates of the $^{42}$Ti($p$,$\gamma$)$^{43}$V with different sources(new, ths8, rath). }

\begin{tabular}{cccc}
\hline\hline
Species & new & ths8 & rath \\
\hline
$^{42}$Ca & 2.01$\times$10$^{-15}$ \textbf{(1.33-2.71)} & 3.55$\times$10$^{-15}$& 3.92$\times$10$^{-15}$\\
				
$^{42}$Sc & 2.76$\times$10$^{-10}$ \textbf{(1.82-3.72)} & 4.86$\times$10$^{-10}$ & 5.38$\times$10$^{-10}$ \\
				
$^{43}$Sc & 5.59$\times$10$^{-14}$ \textbf{(3.69-7.52)} & 1.00$\times$10$^{-13}$ & 1.11$\times$10$^{-13}$\\
				
$^{42}$Ti & 6.21$\times$10$^{-06}$ \textbf{(4.09-8.36)} & 1.09$\times$10$^{-05}$ & 1.21$\times$10$^{-05}$ \\
				
$^{43}$Ti & 1.92$\times$10$^{-09}$ \textbf{(1.27-2.59)} & 3.38$\times$10$^{-09}$ & 3.74$\times$10$^{-09}$\\

$^{45}$Ti & 2.21$\times$10$^{-12}$ \textbf{(2.21-2.21)} & 9.93$\times$10$^{-13}$ & 9.88$\times$10$^{-13}$\\

$^{46}$Ti & 8.35$\times$10$^{-15}$ \textbf{(8.35-8.36)} & 4.19$\times$10$^{-15}$ & 4.16$\times$10$^{-15}$\\

$^{43}$V & 3.65$\times$10$^{-09}$ \textbf{(2.33-4.95)} & 7.24$\times$10$^{-10}$ & 7.39$\times$10$^{-13}$\\

$^{45}$V & 3.23$\times$10$^{-08}$ \textbf{(3.22-3.23)} & 1.45$\times$10$^{-08}$ & 1.44$\times$10$^{-08}$\\

$^{46}$V & 1.34$\times$10$^{-10}$ \textbf{(1.34-1.34)} & 6.71$\times$10$^{-11}$ & 6.68$\times$10$^{-11}$\\

$^{47}$V & 6.00$\times$10$^{-13}$ \textbf{(6.00-6.00)} & 4.82$\times$10$^{-13}$ & 4.80$\times$10$^{-13}$\\

$^{44}$Cr & 1.75$\times$10$^{-06}$ \textbf{(1.12-2.37)} & 3.48$\times$10$^{-07}$ & 3.56$\times$10$^{-10}$\\

$^{45}$Cr & 3.63$\times$10$^{-06}$ \textbf{(3.63-3.63)} & 1.59$\times$10$^{-06}$ & 1.59$\times$10$^{-06}$\\

$^{46}$Cr & 1.70$\times$10$^{-06}$ \textbf{(1.70-1.70)} & 8.51$\times$10$^{-07}$ & 8.49$\times$10$^{-07}$\\

$^{47}$Cr & 8.83$\times$10$^{-10}$ \textbf{(8.83-8.83)} & 5.10$\times$10$^{-10}$ & 5.08$\times$10$^{-10}$\\
\hline
\end{tabular}
\tablefoot{ The data in bold refer to the range of predicted isotopic abundances with the same order of magnitude as the front abundance values when the new forward and reverse rates are allowed to vary within their uncertainty bounds.}	
\label{1}
\end{table}

\section{Conclusion}\label{Sect_Conclu}

Based on a series of improvements on crucial information of the $^{42}$Ti($p$,$\gamma$)$^{43}$V rate calculation, including the spectroscopic factors of the resonances of interest in $^{43}$V, neglected resonance in previous works, and the new nuclear masses of $^{42}$Ti and $^{43}$V, we reevaluate the thermonuclear rate of $^{42}$Ti($p$,$\gamma$)$^{43}$V and its associated uncertainty via a Monte Carlo method. The reverse reaction rate is updated consistently using the new proton separation energy. The present study shows that the previous reaction rates from the statistical model dramatically differ from our new rates. Specifically, the largest differences can be up to two orders of magnitude for the forward reaction and four orders for the reverse reaction over the temperature range of X-ray burst interest in this study. In order to explore the influence of new rates in the rp-process, we perform nucleosynthesis simulations in which three different reaction rates of $^{42}$Ti($p$,$\gamma$)$^{43}$V are used. Compared to the results using ths8 and rath rates, our new rates have no significant impact on the final elemental abundances except for a decrease of 49\% in Calcium and an increase of 128\% in Scandium. In addition, unlike the case using rath rates where the rp-process path bypasses the nucleus $^{43}$V, our results show that the main reaction path will pass through $^{43}$V which is the same as using ths8 rates, but with a larger leakage through $^{43}$V decays.

\begin{acknowledgements}
We thank C.X. Yuan for his shell model calculation on the properties of $^{43}$V excited states. This work was financially supported by the National Key R\&D Program of China Grant no. 2022YFA1603300, the Youth Innovation Promotion Association of Chinese Academy of Sciences under Grant No. 2019406, the Strategic Priority Research Program of Chinese Academy of Sciences Grant No. XDB34020204, and in part by the National Science Foundation under Grant No. OISE-1927130 (IReNA). M.P. acknowledges significant support to NuGrid from the ERC Consolidator Grant (Hungary) funding scheme (Project RADIOSTAR, G.A. n. 724560), from the ChETEC COST Action (CA16117), supported by the European Cooperation in Science and Technology, from the IReNA network supported by NSF AccelNet, from the National Science Foundation (NSF, USA) under grant No. PHY-1430152 (JINA Center for the Evolution of the Elements), from the "Lendulet-2014" Program of the Hungarian Academy of Sciences (Hungary), and from the European Union's Horizon 2020 research and innovation programme (ChETEC-INFRA -- Project no. 101008324). M.P. also acknowledges the access to {\tt viper}, the University of Hull High Performance Computing Facility. the National Natural Science Foundation of China under Grants No. 12205340; J.G.L. acknowledges the support from the National Natural Science Foundation of China under Grants No. 12205340; the Gansu Natural Science Foundation under Grant No. 22JR5RA123;
\end{acknowledgements}

\bibliographystyle{aa} % style aa.bst
\bibliography{ref} % your references Yourfile.bib
% \begin{thebibliography}{}

%   \bibitem[Baker(1966)]{baker} Baker, N. 1966,
%       in Stellar Evolution,
%       ed.\ R. F. Stein,\& A. G. W. Cameron
%       (Plenum, New York) 333

%    \bibitem[Balluch(1988)]{balluch} Balluch, M. 1988,
%       A\&A, 200, 58

%    \bibitem[Cox(1980)]{cox} Cox, J. P. 1980,
%       Theory of Stellar Pulsation
%       (Princeton University Press, Princeton) 165

%    \bibitem[Cox(1969)]{cox69} Cox, A. N.,\& Stewart, J. N. 1969,
%       Academia Nauk, Scientific Information 15, 1

%    \bibitem[Mizuno(1980)]{mizuno} Mizuno H. 1980,
%       Prog. Theor. Phys., 64, 544
   
%    \bibitem[Tscharnuter(1987)]{tscharnuter} Tscharnuter W. M. 1987,
%       A\&A, 188, 55
  
%    \bibitem[Terlevich(1992)]{terlevich} Terlevich, R. 1992, in ASP Conf. Ser. 31, 
%       Relationships between Active Galactic Nuclei and Starburst Galaxies, 
%       ed. A. V. Filippenko, 13

%    \bibitem[Yorke(1980a)]{yorke80a} Yorke, H. W. 1980a,
%       A\&A, 86, 286

%    \bibitem[Zheng(1997)]{zheng} Zheng, W., Davidsen, A. F., Tytler, D. \& Kriss, G. A.
%       1997, preprint
% \end{thebibliography}

\end{document}